\documentclass[preprint]{aastex}
\usepackage{amssymb,amsmath,graphicx,epsfig}
\newcommand{\sout}[1]{}   
\newcommand{\pout}[1]{}   
\def\gsim{\\raise 3pt\hbox{$\rangle$}\kern -8.5pt\raise -2pt\hbox{$\sim$}\ }
\def\lsim{\\raise 3pt\hbox{$\langle$}\kern -8.5pt\raise -2pt\hbox{$\sim$}\ }

\def\gsim{\ \raise 3pt \hbox{$\rangle$} \kern -8.5pt \raise -2pt \hbox{$\sim$}\ }
\def\lsim{\ \raise 3pt \hbox{$\langle$} \kern -8.5pt \raise -2pt \hbox{$\sim$}\ }
\shorttitle{3D Structure of Gyroresonance Sources vs Magnetic Extrapolations} \shortauthors{Nita et al.}

\begin{document}
\title{3D Structure of Microwave Sources from Solar Rotation Stereoscopy vs Magnetic Extrapolations}
\author{Gelu M. Nita\altaffilmark{1},
Gregory D. Fleishman\altaffilmark{1,2}, Ju Jing\altaffilmark{3}, Sergey V. Lesovoi\altaffilmark{4}, Vladimir M. Bogod\altaffilmark{5}, Leonid V. Yasnov\altaffilmark{6}, Haimin Wang\altaffilmark{3}, and Dale E. Gary\altaffilmark{1}}
\altaffiltext{1}{Center For Solar-Terrestrial Research, New Jersey Institute of Technology, Newark, NJ 07102}
\altaffiltext{2}{Ioffe Institute, St. Petersburg 194021, Russia}
\altaffiltext{3}{Space Weather Research Lab, New Jersey Institute of Technology, Newark, NJ 07102}
\altaffiltext{4}{Institute of Solar-Terrestrial Physics, Irkutsk 664033, Russia}
\altaffiltext{5}{Special Astrophysical Observatory (SAO), St. Petersburg 196140, Russia}
\altaffiltext{6}{Radiophysics Research Institute of St. Petersburg State University, St. Petersburg 198504, Russia}

\begin{abstract}
We use rotation stereoscopy to estimate the height of a steady-state solar feature relative to the photosphere, based on its apparent motion in the image plane recorded over several days of observation. The stereoscopy algorithm is adapted to work with either one- or two-dimensional data (i.e. from images or from observations that record the projected position of the source along an arbitrary axis). The accuracy of the algorithm is tested on simulated data, and then the algorithm is used to estimate the coronal radio source heights associated with the active region NOAA 10956, based on multifrequency imaging data over 7 days from the Siberian Solar Radio Telescope near 5.7~GHz, the Nobeyama Radio Heliograph at 17~GHz, as well as one-dimensional scans at multiple frequencies spanning the 5.98--15.95~GHz frequency range from the RATAN-600 instrument. The gyroresonance emission mechanism, which is sensitive to the coronal magnetic field strength, is applied to convert the estimated radio source heights at various frequencies, $h(f)$, to information about magnetic field vs. height $B(h)$, and the results are compared to a magnetic field extrapolation derived from photospheric magnetic field observations obtained by Hinode and MDI. We found that the gyroresonant emission comes from the heights exceeding location of the third gyrolayer irrespectively on the magnetic extrapolation method; implications of this finding for the coronal magnetography and coronal plasma physics are discussed.
\end{abstract}


\section{Introduction}
\label{S_intro}

Most coronal physics, including various kinds of solar activity and space weather drivers, depends critically on the coronal magnetic field, which, however, is very difficult to reliably measure. Some clues about the magnetic loop structure can be derived from various EUV line emission produced by relatively dense plasma at certain temperature ranges. The corresponding loop-like 3D structures of the EUV brightness (commonly called EUV loops) do not necessarily trace the magnetic field lines precisely because the EUV brightness depends on a combination of the emission measure (EM), temperature and isotopic abundance of the plasma, thus, the \emph{brightness loop} generally deviates from a \emph{magnetic field line} \citep[e.g.,][]{Mok_etal_2008}.

A current state of the art in the analysis of coronal magnetic field is magnetic field extrapolations from the photospheric level, where either line-of-sight or even vector magnetic field measurements are available from both ground- and space-based solar telescopes. There is no unique way of extrapolating the magnetic field from the forced photospheric boundary up to the force-free corona. Therefore, a number of different techniques have been developed including potential field (PF), linear or nonlinear force-free field (NLFFF) extrapolations, or fits to multiple dipole field models \citep{Aschw}. Various tests for those extrapolations bring controversial results \citep[e.g.,][]{deRosa_etal_2009}.  In particular, comparison of the magnetic field lines with the 3D structure of the EUV loops observed by STEREO high in the corona has led \citet{deRosa_etal_2009} to conclude that a potential extrapolation works better than the corresponding NLFFF extrapolation, while  \citet{Aschw} concluded that a field composed of a few subphotospheric dipoles works even better. On the other hand, it is well established that solar activity requires free magnetic energy to be released in the corona, which implies that the coronal field must somehow deviate from the potential one \citep[e.g.,][]{Devore_2005, Fl_etal_2011} at some regions, perhaps at the coronal base. It is recognized that magnetograms based on chromospheric (rather than photospheric) observations could provide a more suitable boundary condition for the extrapolations. To choose among and improve upon these various magnetic field models or extrapolation techniques requires verification by comparison with independently measured coronal magnetic field.

In contrast to EUV and X-ray emission, which are not directly sensitive to the magnetic field vector at the source, gyrosynchrotron and gyroresonance (GR) radio emissions depend on the magnetic field explicitly, so measurable characteristics of the radio emissions contain direct information on the magnetic field. The gyrosynchrotron emission requires fast electrons to be produced in large numbers to dominate the radio spectrum, and, thus, is only usable during solar flares \citep{Fl_etal_2009}, while GR emission is produced by the steady-state thermal plasma and so is appropriate for routine probing of the magnetic field \citep[see, e.g.,][for a review]{Lee_2007}. Note that although a non-flaring microwave continuum component, distinct from the GR one, is often observed along with GR emission, which implies the presence of nonthermal electrons in the ARs outside of flaring times  \citep{Akhmedov_etal_1986, Kaltman_etal_1998}, its use as a diagnostic of coronal magnetic fields has not been developed. Unfortunately, the GR emission does not directly provide height information for the GR sources; instead, a single imaging spectroscopy observation of GR sources in an active region provides the plasma temperature as a function of the emission frequency, which, being a small integer multiple of the gyrofrequency, is directly proportional to the magnetic field.

Therefore, to derive the 3D structure of the coronal magnetic field requires additional height information. One way of getting the height information is to rely on a magnetic field model and fit the radio contours to the corresponding isogauss surfaces. An apparent disadvantage of this approach is that it requires a reliable model and so cannot be used to verify models. This paper attempts a different approach, proposed by \cite{Aliss_1984, RS1, RS2}, namely, to use solar rotation to estimate the 3D locations of the GR sources above a relatively steady active region. To do so we develop a new optimization algorithm employing solar rotation stereoscopy, test it on simulated data, and apply it to multi-instrument imaging observations performed by the Siberian Solar Radio Telescope \citep[SSRT,][]{SSRT}, RATAN-600 \citep{Bogod_etal_1999}, and the Nobeyama Radio Heliograph \citep[NoRH,][]{NORH} . We then check the radio measurements against the magnetic extrapolations and discuss fundamental implications of our findings.

\section{Algorithm Description}
Our optimization algorithm assumes a model in which a steady state structure moves with the photospheric sunspot, at a fixed height $h$, while allowing a constant displacement defined by the yet unknown parameters $\delta\lambda$ and $\delta\varphi$ relative to the assumed constant AR latitude $\lambda$ and the changing AR longitude $\varphi=\varphi_0+\omega(t-t_0)$, where $\varphi_0$ is the initial AR longitude at some reference time $t=t_0$. Besides the two free parameters $\delta\lambda$ and $\delta\varphi$, the optimization routine tunes the parameter $\rho=R/R_{Sun}$, which is the ratio between a yet unknown rotation radius of the source and the apparent solar radius expressed in arcsec. Using $\rho$ rather than $R$ makes the result independent of the time varying Earth-Sun distance. The trial heliospheric coordinates $(R=\rho R_{Sun},\varphi+\delta\varphi, \lambda+\delta\lambda)$, are converted to heliocentric coordinates $(\widetilde{x}_i,\widetilde{y}_i)$ and are combined with the observed heliocentric coordinates $(x_i,y_i)$ of the tracked feature, in order to minimize the two-dimensional objective function
\begin{eqnarray}
\label{obj1}
&&\sigma_{2D}^2(\rho,\delta\varphi,\delta\lambda)=\\\nonumber
&&\frac{1}{2N-n}\sum_{i=1}^N\{[x_i-\widetilde{x}_i(\rho,\delta\varphi,\delta\lambda)]^2+[y_i-\widetilde{y}_i(\rho,\delta\varphi,\delta\lambda)]^2\},
\end{eqnarray}
representing the data--model squared residual normalized by the number of degrees of freedom $2N-n$, with the number of free model parameters $n$, which may range from $1$ to $3$, depending on the number of free parameters allowed. Though the minimized residual, measured in arcsec, is determined by the combined uncertainties of all three optimized parameters, nevertheless it puts an upper limit on the uncertainty affecting the derived source height, and so provides an estimate of the error affecting this parameter, which we express as $h=(\rho-1)R_{Sun}\pm\sigma_{2D}$.

To accommodate the option of using this optimization algorithm with data that result from  one-dimensional observations that give only the projection of the source position along an axis perpendicular on the local terrestrial meridian, which makes an angle $P$ with solar rotational axis, we write the projected coordinate as
\begin{eqnarray}
\label{xp}
\xi=x \cos(P)-y\sin(P)
\end{eqnarray}
and define an alternative one-dimensional objective function
\begin{eqnarray}
\label{obj2}
&&\sigma_{1D}^2(\rho,\delta\varphi,\delta\lambda)=\frac{1}{N-n}\sum_{i=1}^N[\xi_i-\widetilde{\xi}_i(\rho,\delta\varphi,\delta\lambda)]^2,
\end{eqnarray}
which, after minimization, provides the height estimate $h=(\rho-1)R_{Sun}\pm\sigma_{1D}$.

Basically, this variant of the algorithm is similar to the method used by \citet{RATAN} to estimate radio source heights from RATAN-600 data but, compared with their method,  allows two more degrees of freedom, i.e $\delta\varphi$ and $\delta\lambda$.

\section{Algorithm Performance Test}
\begin{figure}
\epsscale{1.1}\plotone{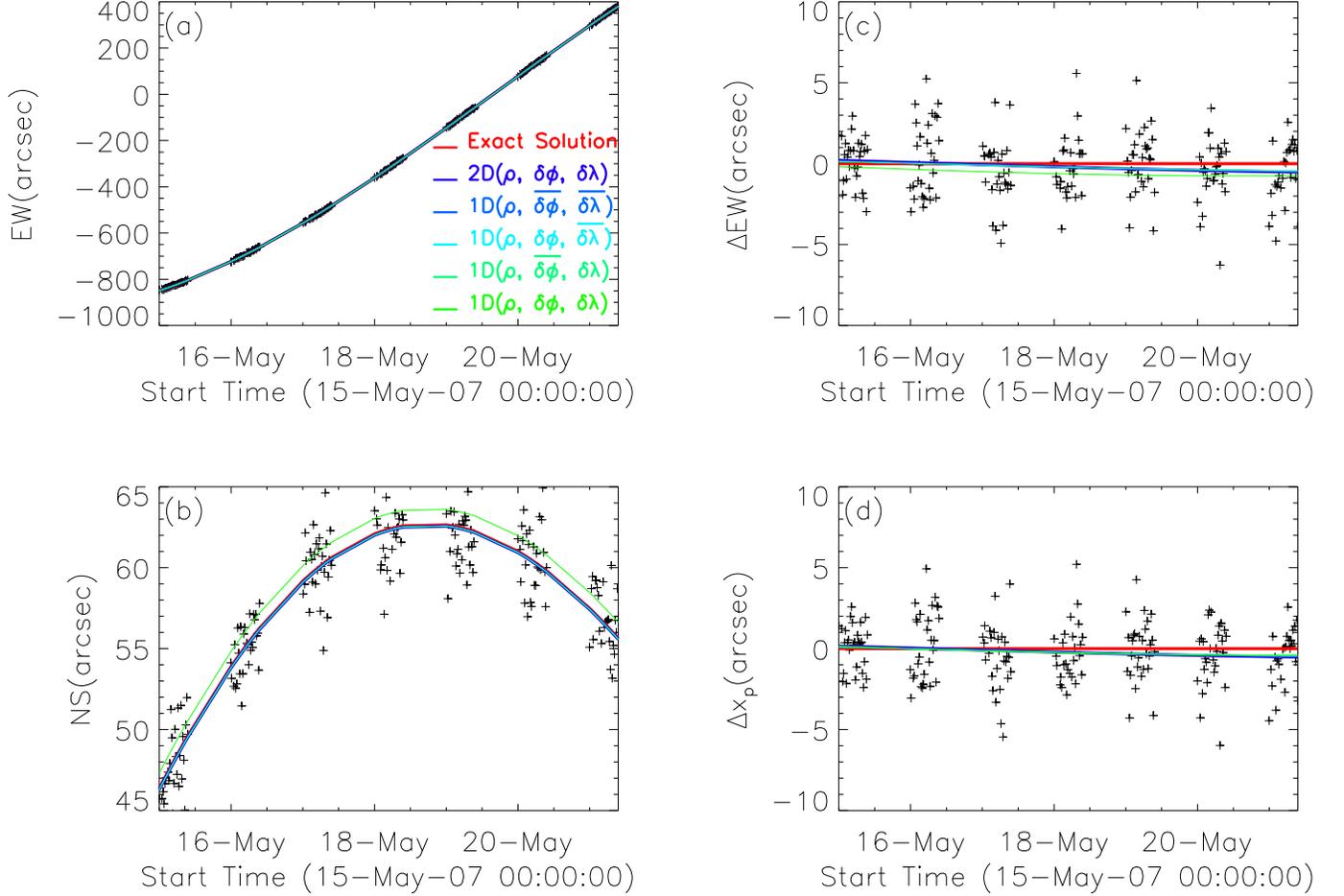}\caption{\label{sim} The results of the 2D and 1D optimization routines for simulated data, as explained in the main text. In the four cases corresponding to the 1D optimization, the overlined arguments in panel (a) were kept fixed to the values returned by the 2D optimization routine.}
\end{figure}

In order to test the performance of the algorithm, we used numerically simulated data according to the model we have adopted. Our test consisted of a dataset simulating the heliocentric positions of a steady state structure, assumed to have been located above the AR NOAA 10956, and recorded over 7 days between 15-May-2007 and 21-May-2007 with a 20-minute time resolution during 8 h of daily observations. The choice of position and time range for our simulated data is motivated by the analysis of actual data presented in the next section, which refers to this particular AR. The simulated feature is characterized by the exact parameters $h/R_{Sun}=1\%$, $\delta\varphi=-0.5^{\circ}$, and $\delta\lambda= 0.5^{\circ}$. However, to account for instrumental pointing errors, we have added realistic gaussian random noise to each data point, resulting in an overall residual, computed according to equation (\ref{obj1}), of $\sigma=1.930"$, i.e. $\sigma/R_{Sun}^*=0.203\%$, where $R_{Sun}^*=948.8''$ represents the average apparent solar radius for the considered time interval. The simulated EW and NS positions are plotted in Figure \ref{sim} (plus signs), panels (a) and (b), respectively. Panel (c) shows the residuals---the same data as in panel (a), but after the exact position has been subtracted from the data. This allows a better evaluation of the errors affecting each data point. Panel (d) shows the difference between the simulated projected position, $\xi$, computed according to equation (\ref{xp}), and the exact solution for the time evolution of this projection. On each panel we show, as solid lines, the exact solution (red), the results of the three-parameter 2D optimization (dark blue), as well as the solution given by the 1D optimization routine in four cases corresponding to different degrees of freedom resulting from fixing none, one, or both of the relative angular displacement parameters to the corresponding values returned by the 2D optimization routine.

The results displayed in Figure \ref{sim} show that the evolution of the exact solution is reconstructed with reasonable accuracy by both the 2D and 1D optimization routines.
The numerical results of this test are listed in Table \ref{test}, where the last two columns indicate the overall residual computed according to equations (\ref{obj1}) or (\ref{obj2}), as absolute and relative values. These numerical results show that the 2D optimization routine is able to estimate all three model parameters within the true error level affecting the simulated data set, which is also accurately estimated. Interestingly, the 1D optimization routine gives more accurate estimates of the true height, but at the cost of less accurate displacement estimates and underestimation of the  error level. However, in all analyzed cases the height is correctly estimated within the true error level, which we consider to be a clear indication of the robustness and reliability of this approach.

\section{Data Analysis}

We employed the optimization method presented in the previous section to estimate the radio source heights at various frequencies associated with NOAA AR 10956, assumed to have no significant systematic structural changes over the time interval considered other than, perhaps,  quasiperiodic variations of the temperature due to thermal instability \citep{Mok_etal_2008} or short-term dynamics due to fluctuations in photospheric boundary conditions. The apparent centroid position of the GR source from a realistic AR was shown \citep{Gelfreikh_Lubyshev_1979} to depend on the AR longitude due to the viewing angle effect. The available data are insufficient to fully address this effect; however, we make an assessment of it later. The data set consists of data from the Siberian Solar Radio Telescope (SSRT) at 5.7~GHz, the Nobeyama Radio Heliograph (NoRH) at 17~GHz, as well as one-dimensional scans at multiple frequencies spanning the 5.98--15.95~GHz frequency range from the RATAN-600 instrument. As the first step of our analysis, we have used photospheric magnetic field maps inferred from MDI data to track the movement of AR 10956 on the solar surface from 13-May-2007 to 23-May-2007. The heliocentric coordinates of the AR were then converted into heliospheric, and the longitudinal rotation speed of $\omega=(1.5501\pm 0.0003)\times10^{-4}$~deg/sec has been determined by a linear fit of the longitudinal coordinates. We found that the latitude of the AR inferred from the MDI maps, $\lambda=(0.99\pm0.11)^\circ$N, was consistent with the general solar differential rotation formula, which gives a rotation speed of $\omega=(1.54671\pm0.00002)\times10^{-4}$~deg/sec for the same latitude range.

\subsection{Radio  Position Estimation From Imaging Data}
The inferred rotation speed and AR latitude were further used as known input parameters for the 2D optimization algorithm used to infer the radio source heights from the SSRT and NoRH maps at 5.7~GHz and 17~GHz, respectively.

\begin{figure}
\epsscale{1.1}\plotone{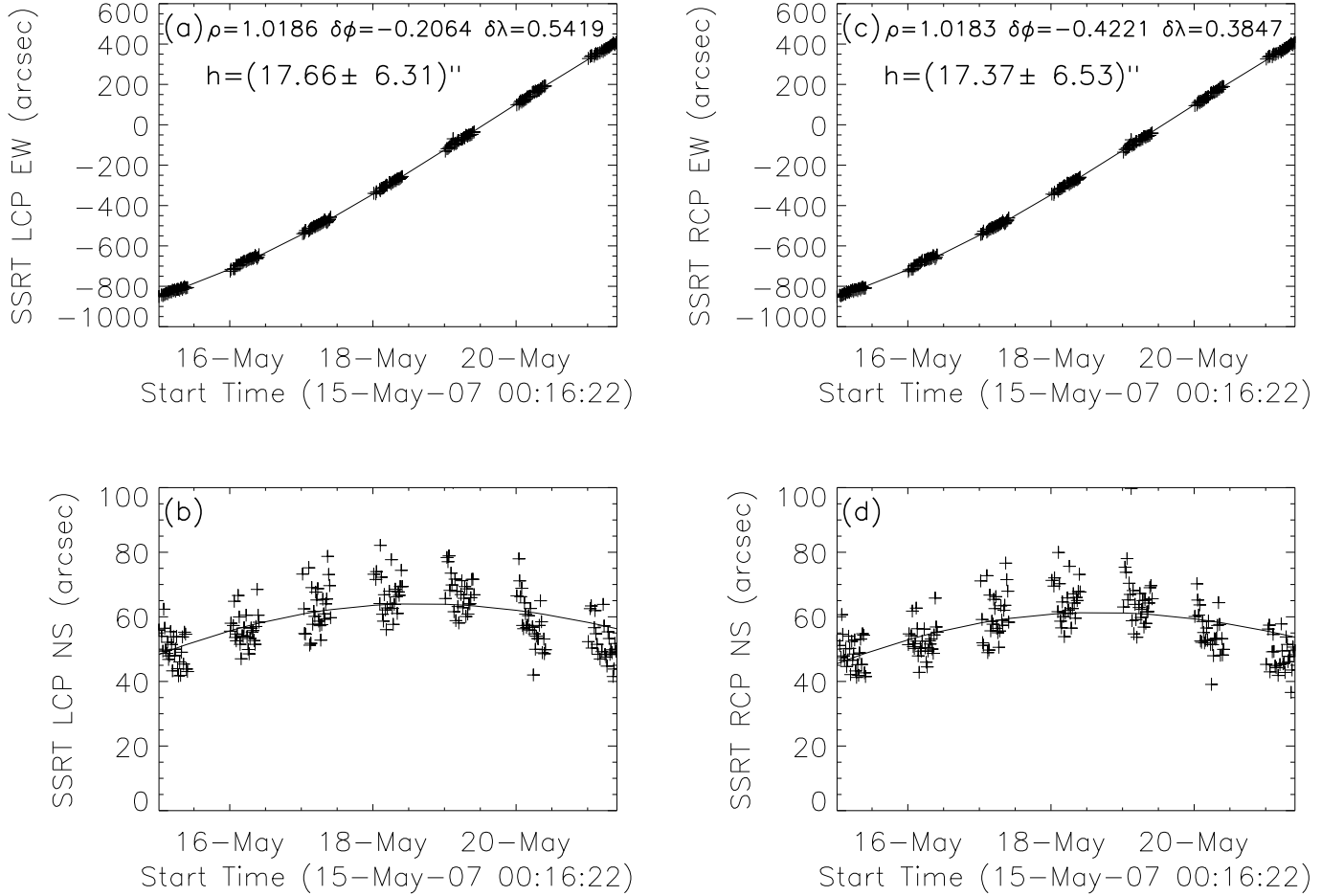}\caption{\label{ssrt_2dall} Optimization results based on SSRT radio maps at 5.7~GHz from 15-May-2007 to 21-May-2007 for the LCP (left column) and RCP (right column) circular polarizations. Top row: data (plus signs) and the optimized solution (solid line) corresponding to the observed EW heliocentric locations of the radio LCP and RCP centroids. Bottom row: data (plus signs) and the optimized solution (solid line) corresponding to the observed NS heliocentric locations of the radio LCP and RCP centroids. The optimized parameters and the source height estimation for the LCP and RCP polarizations are displayed in the figure insets of (a) and (c) panels, respectively. Since no noticeable structural change has been observed in the radio maps, the systematic NS scatter of the centroid locations is attributed to the instrumental pointing errors/quasiperiodic AR dynamics.}
\end{figure}

Figure \ref{ssrt_2dall} displays the 2D optimization results based on the SSRT radio maps at 5.7~GHz from 15-May-2007 to 21-May-2007 for the LCP (left column) and RCP (right column) circular polarizations. The data (plus signs) and the optimized solution (solid line) corresponding to the observed EW heliocentric locations of the radio LCP and RCP centroids are shown on the top row, while the bottom row displays data and solution corresponding to the observed NS heliocentric locations of the radio LCP and RCP centroids. Since no noticeable structural changes have been observed in the SSRT radio maps, the systematic NS scatter of the centroid locations is attributed to the instrumental pointing errors and/or relatively short-term active region dynamics.
We did not find any dissimilarities between the fits and residuals for the simulated and real data, respectively, which we regard as evidence for a reasonable constancy of the apparent position of the radio centroid relative to the AR and, thus, justifies that the viewing angle \citep{Gelfreikh_Lubyshev_1979} has little effect, within the uncertainties, in our case.
The optimized solution provides the radii and displacements relative to the heliocentric AR coordinates for the LCP and RCP polarizations, i.e. $\left(\rho=1.0186, \delta\phi=-0.2064,\delta\lambda=0.5419\right)$ and $\left(\rho=1.0183, \delta\phi=-0.4221,\delta\lambda=0.3847\right)$, respectively. The estimated relative heights of the radio centroids have been converted to the apparent heights $h=\left(17.66\pm 6.31\right)''$, and $h=\left(17.37\pm 6.53\right)''$, respectively, corresponding to $R_{Sun}=948.1''$, which was the earth-view apparent solar radius as observed on 18-May-2007 06:06:20 UT, the time of the Hinode magnetograms used to derive the extrapolated magnetic field models discussed later.

\begin{figure}
\epsscale{1.1}\plotone{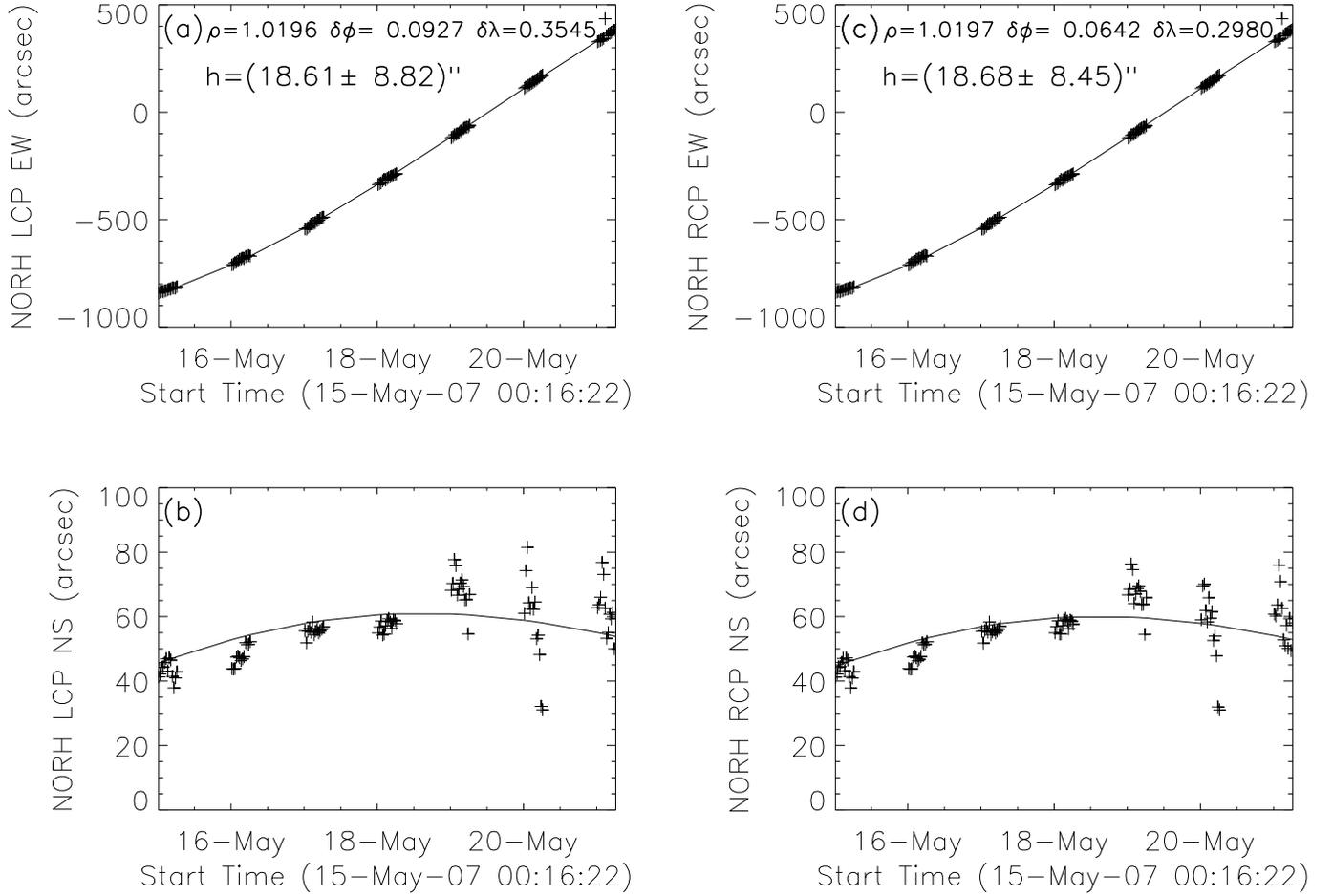}\caption{\label{norh_2dssrtmatch} 2D optimization results for the LCP (left column) and RCP (right column) circular polarizations based on NoRH radio maps at 17~GHz spanning the same time interval as the SSRT maps used to produce the results displayed in Figure \ref{ssrt_2dall}. The structure of this multipanel figure is the same as in Figure \ref{ssrt_2dall}. A larger scatter of the NS locations may be observed starting with 19-May-2007, which is due to the observed change of the 17~GHz radio source from a unipolar to a bipolar structure.}
\end{figure}

\begin{figure}
\epsscale{1.1}\plotone{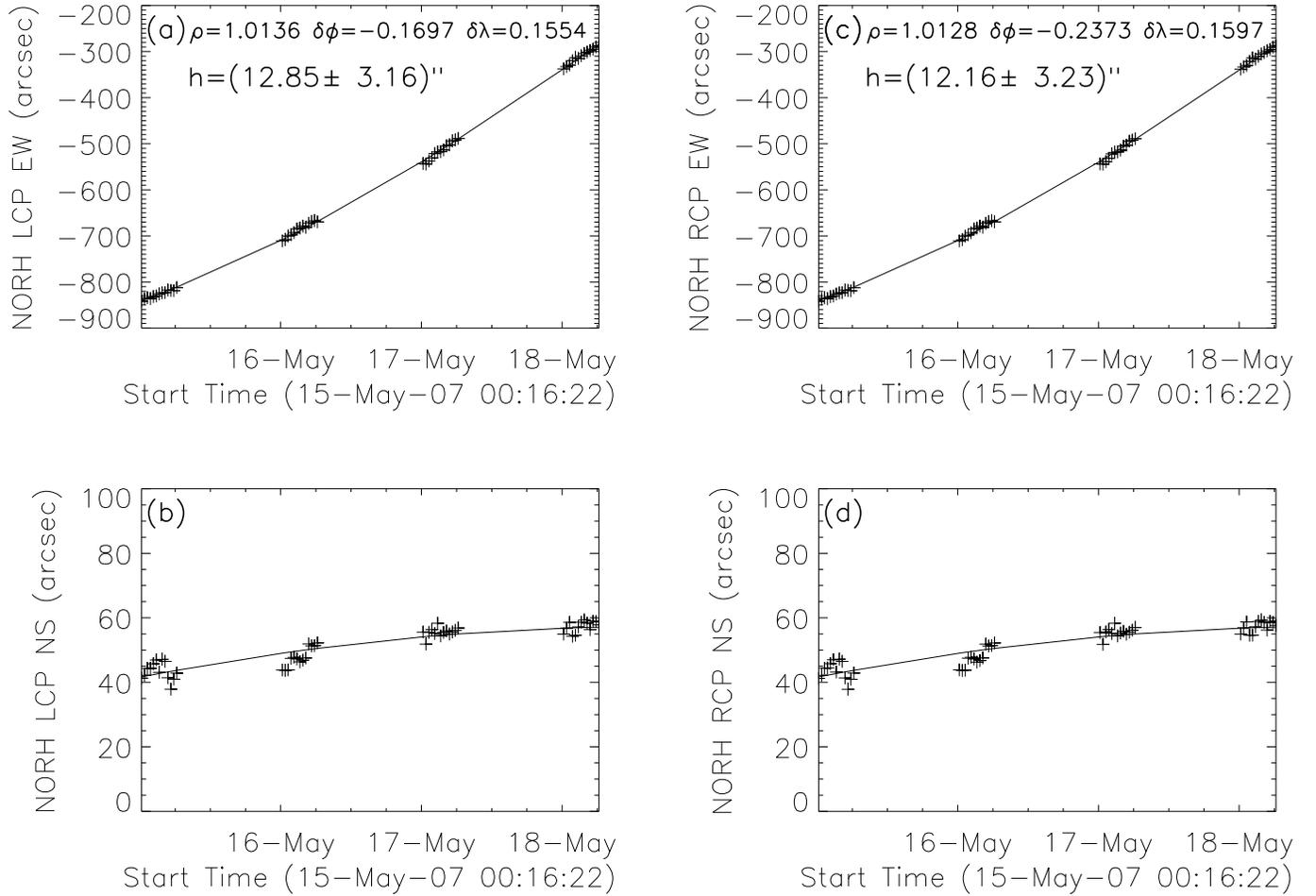}\caption{\label{norh_2dshort} 2D optimization results for the NoRH radio source at 17~GHz based on a time range restricted to the interval in which the source structure is unipolar. The structure of this multipanel figure is the same as in Figure \ref{ssrt_2dall}. Lower source heights and smaller height uncertainties than in Figure \ref{norh_2dssrtmatch} are derived from this restricted time range.}
\end{figure}

Figure \ref{norh_2dssrtmatch} presents the 2D optimization results based on the NoRH radio maps at 17~GHz covering the same time interval as SSRT. At first glance, the estimated LCP and RCP source heights at 17~GHz, $h=(18.61\pm 8.82)''$ and $h=(18.68\pm 8.45)''$, respectively, are not consistent with our expectation for the higher frequency radio sources to be located at lower altitudes above the photosphere, although one may argue that the larger error bars affecting these estimates do not allow for a conclusive comparison of the SSRT and NoRH heights. However, note the larger scatter in the NS heliocentric positions of the 17~GHz radio sources mapped in the last 3 days of the observed period, which may be responsible for the large error bars in the NoRH height estimation. Indeed, an examination of the morphology of the NoRH radio maps, not detailed here, showed that the 17~GHz source began with a unipolar structure, but developed a bipolar structure starting with 19-May-2007 observations, which makes the subsequent data inconsistent with the assumed model of a steady state structure, upon which the development of our optimization algorithm is based. Consequently, we present in Figure \ref{norh_2dshort} the 17~GHz radio source height estimations based solely on the data recorded by NoRH from 15-May-2007 to 18-May-2007. The NoRH 17~GHz estimates obtained from the restricted data set, i.e. $h=(12.85\pm 3.16)''$ for LCP, and $h=(12.16\pm3.23)''$ for RCP, have smaller error bars, and correspond to lower altitudes than the ones corresponding to the lower frequency estimates derived from the full SSRT data set.

\begin{figure}
\epsscale{1.1}\plotone{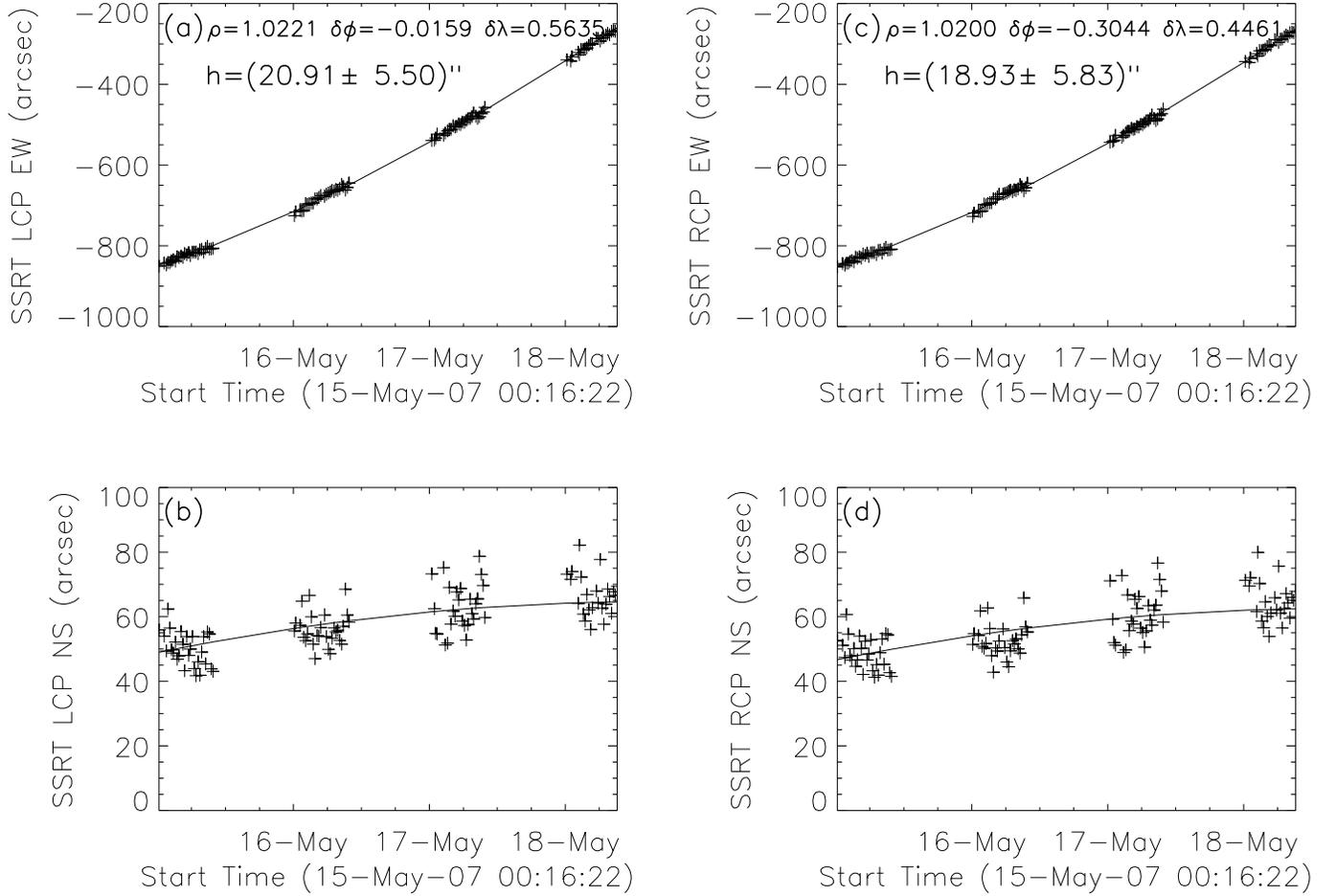}\caption{\label{ssrt_2dshort} Optimization results based on SSRT radio maps at 5.7~GHz from 15-May-2007 to 18-May-2007, the time range showing a unipolar source at 17~GHz. The structure of this multipanel figure is the same as in Figure \ref{ssrt_2dall}. The heights derived from this restricted time range are larger than those shown in Figure \ref{ssrt_2dall}, which may be a consequence of selecting only radio sources located on one side of the solar meridian as opposed to a collection of symmetrical viewing angles.}
\end{figure}

However, to allow for a fair comparison of NoRH and SSRT results, we present in Figure \ref{ssrt_2dshort} the SSRT estimates derived from the same restricted time interval as used for the NoRH data set. The 5.7~GHz height estimates derived from this restricted time range, i.e. $h=(20.91\pm 5.50)''$ for LCP, and $h=(18.93\pm5.83)''$ for RCP, are slightly larger than those shown in Figure \ref{ssrt_2dall}, which may be a consequence of selecting only radio sources located on one side of the solar meridian as opposed to a collection of symmetrical viewing angles. One may expect that the \citet{Gelfreikh_Lubyshev_1979} viewing angle effect will be most prominent for this asymmetric collection of the data
thus, having the estimated heights change by only $1.5-3''$ (within the estimated uncertainty of $5-6''$) for symmetric and asymmetric data sampling confirms that the viewing angle effect plays only a minor role in our study, if any. The results displayed in Figure \ref{ssrt_2dshort} suggest that the heights derived from data displayed in Figure \ref{norh_2dshort} may be also biased toward larger heights than those that would have been obtained if the unipolar source structure had survived for the full time range of the 17~GHz NoRH observations.

\begin{figure}
\epsscale{0.7}\plotone{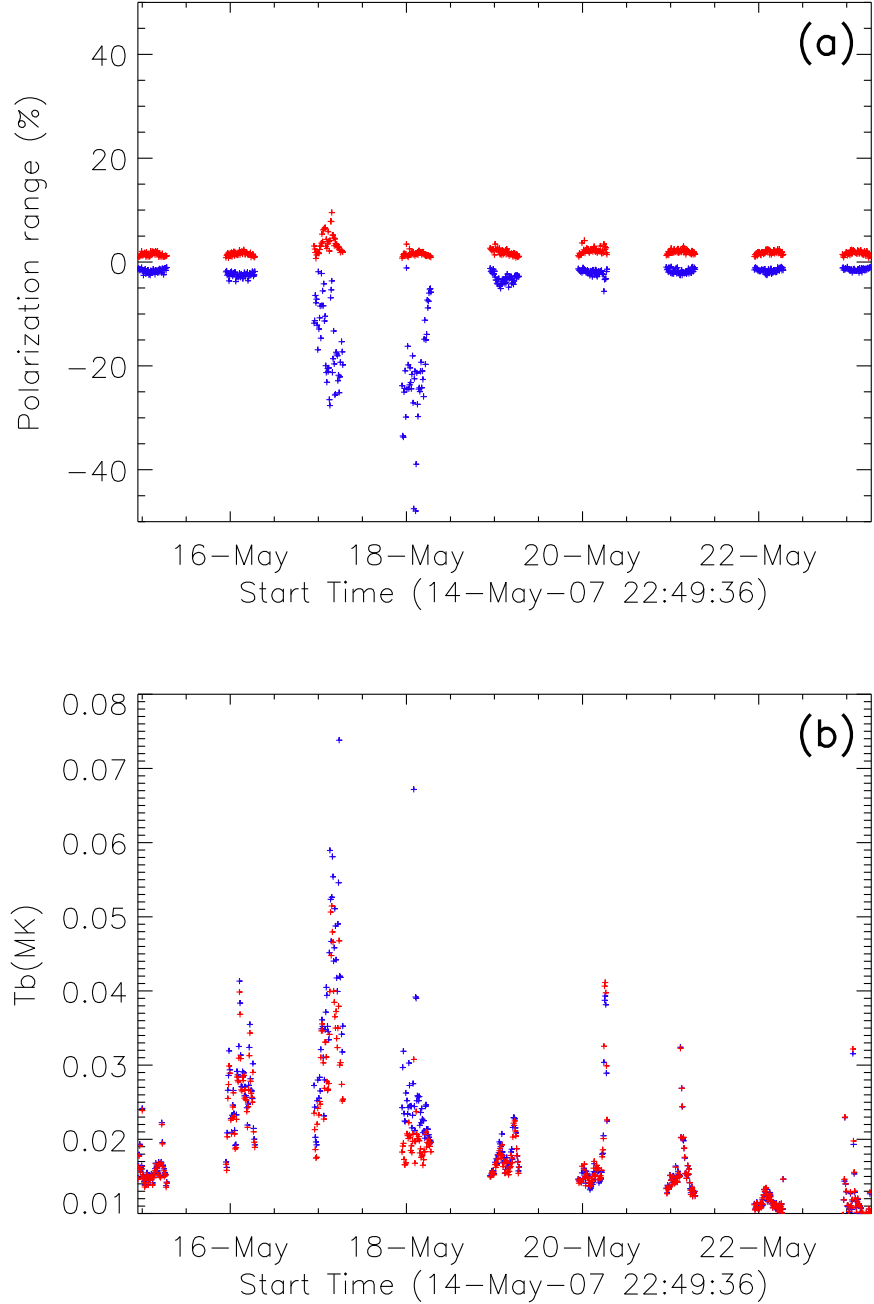}\caption{\label{norh_pol}(a) The extremes of the negative (LCP--blue) and positive (RCP--red) polarizations in the 17GHz NoRH maps spanning several days  of observation. Relatively high degrees of LCP polarization are observed on the two days preceding May 19 2007. (b) The maximum LCP(blue) and RCP(red) brightness temperatures, $T_b$(MK), at 17GHz corresponding to the same data set. The adopted convention for the brightness temperature is $T_b(I)=T_b(RCP) + T_b(LCP)$. Similar to the results shown in panel (a), the evolution of the brightness temperatures also displays noticeable variations during the time interval considered.}
\end{figure}

Although no morphological change was apparent before May 19, the brightness temperature and the degree of polarization at 17~GHz were found to evolve, as shown in Figure \ref{norh_pol}. The brightness temperature (in Stokes I) varies between 0.02 and 0.12~MK, reaching the peak values on May 17, while the degree of circular polarization grows to 10--50\% on May 17-18, reaching the peak values of $\sim50\%$ on May 18, indicative of a dominant contribution from gyro emission at 17~GHz during those two days at least. We note that on May 19 some AR restructuring low in the corona, revealing itself in NoRH data as morphological changes in the images, a drop in the degree of polarization, and a drop in brightness temperature, was followed after the end of the NoRH observing day by an eruption that occurred on May 19, 12:48~UT \citep{Li_etal_2008}.

\subsection{Radio Source Position Estimation From RATAN-600 1D Scans}
In the next stage of our analysis, we use one-dimensional scans at multiple frequencies from the RATAN-600 instrument to estimate the radio source heights for both circular polarizations. The RATAN-600 data consists of a set of 53 frequencies spanning the 5.98--15.95~GHz frequency range. To minimize the pointing errors, the data were taken only once a day, when the Sun was on the local meridian. Since the significantly lower time resolution relative to those of the SSRT and NoRH instruments resulted in unstable 1D optimization estimates when all three parameters were allowed to vary, we used the 2D optimization results to add reasonable constraints to the 1D optimization process. Even so, the residuals corresponding to the 1D optimization solution were significantly larger than in the case of the 2D optimization, which we consider to be the combined result of the reduced information contained by the 1D data and the significantly smaller number of available data points.

\begin{figure}
\epsscale{0.7}\plotone{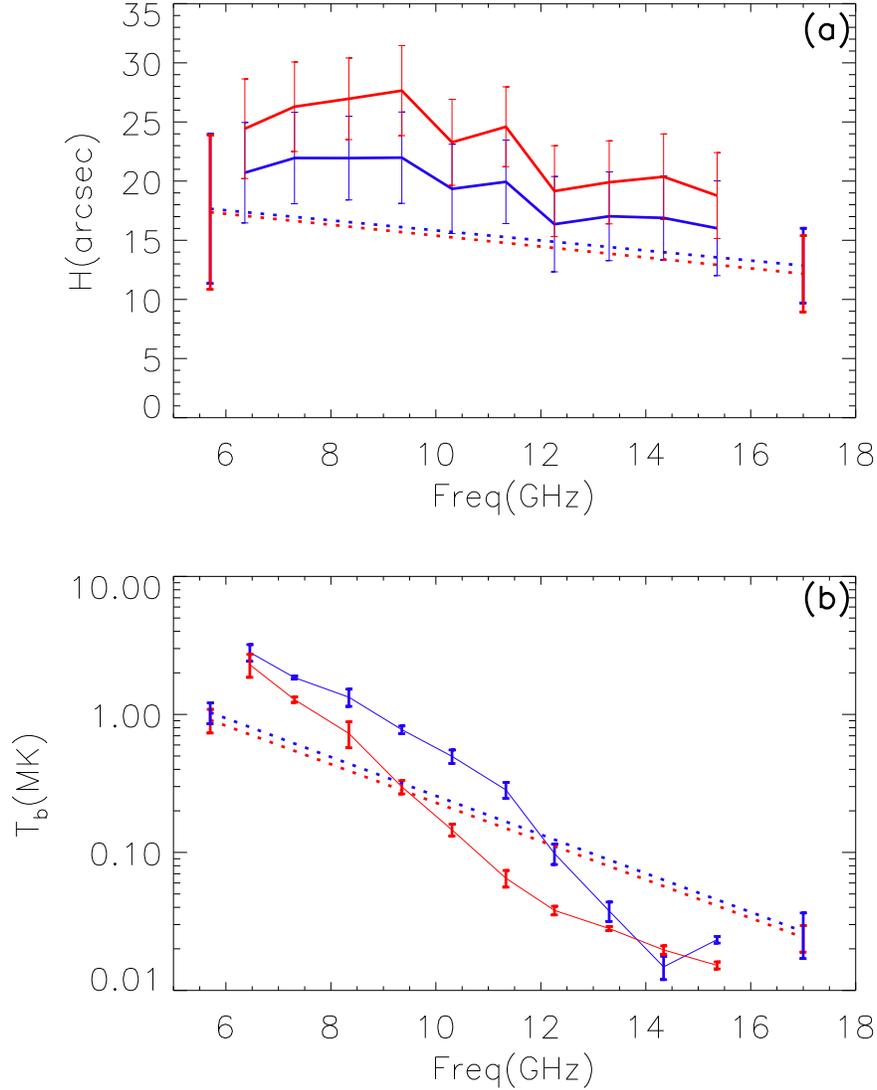}\caption{\label{h} Panel~(a): The LCP (blue) and RCP (red) radio source height estimates, and their corresponding uncertainties, as derived from SSRT (5.7~GHz), RATAN-600 (5.98--15.95~GHz), and NoRH (17~GHz) data sets. The SSRT and NoRH height estimates, which are connected by dotted lines, were obtained using the 2D optimization routine with three degrees of freedom. The RATAN-600 estimates were obtained using the 1D optimization routine with one degree of freedom by fixing the RATAN-600 angular displacements of the LCP and RCP sources to frequency dependent positions obtained by linear interpolation between the corresponding extremes at 5.7~GHz and 17~GHz given by the SSRT and NORH 2D estimates, respectively, and letting only the RATAN heights run as free parameters. The RATAN-600 single frequency estimates were averaged over ten adjacent 1~GHz frequency ranges. Panel~(b): Brightness temperature vs. frequency, derived from SSRT, RATAN-600, and NoRH data of May-18-2007. The SSRT and NORH $T_{b}$ estimates, and their corresponding standard deviations, where obtained by averaging multiple observations during the same day, while the single frequency RATAN-600 estimates were averaged over the same frequency ranges as in panel~(a).}
\end{figure}

Figure \ref{h}a shows the RATAN-600 estimates that we found to be most consistent with the SSRT and NoRH estimates. The results displayed in this figure were obtained by fixing the RATAN-600 angular displacements of the LCP and RCP sources to frequency dependent positions obtained by linear interpolation between the corresponding extremes at 5.7~GHz and 17~GHz given by the SSRT and NORH 2D estimates listed in the insets of Figures \ref{ssrt_2dall} and \ref{norh_2dshort}, respectively, and letting only the RATAN heights run as free parameters.To further reduce the uncertainties of the single frequency RATAN-600 estimates, they were averaged over ten adjacent 1~GHz frequency ranges.

Therefore, in the limit of the estimated uncertainties, we may conclude that the results of the 2D and 1D optimization routines are consistent with a model in which the radio source heights decrease with increasing frequency, which is expected for the AR radio emission formed by both gyroresonance and free-free emission mechanisms.

To help assess over what range of frequencies the GR emission mechanism dominates the observed multifrequency emission, we show the derived brightness temperatures ($T_{\rm b}$) in Figure \ref{h}b.  The brightness temperatures for the SSRT and NoRH points are obtained directly from the images, while those from the RATAN-600 data are derived assuming the sources are circular, with the same N-S extent as the measured E-W extent in the 1D scans.  The plot is consistent with GR emission with $T_{\rm b} > 0.1$~MK up to about 12 GHz, being highly polarized above about 8 GHz.  The lower mean $T_{\rm b}$ above 12~GHz is indicative of free-free emission primarily from the chromosphere and transition region. Nevertheless, similar to the 17~GHz case, we suspect some contribution from gyro emission to play a role in the RATAN-600 observations above 12~GHz.

\section{Potential and Nonlinear Force-Free  Modeling of Coronal Magnetic Fields}

We further use the estimated radio source heights to perform a consistency check of the extrapolated magnetic field structure derived by NLFFF and PF extrapolations.

The Spectro-Polarimeter (SP) instrument of the Solar Optical
Telescope (SOT) on board \emph{Hinode} obtains Stokes profiles of
two magnetically sensitive Fe lines at 630.15 and 630.25 nm with a
sampling of 21.6 m\AA. The \emph{Hinode}/SOT-SP scan of this
active region (NOAA 10956) begins at 06:06 UT, May 18, 2007 and
takes $\sim$30 min to complete. The polarization spectra were
inverted to create the photospheric vector magnetogram using an
Unno-Rachkovsky inversion based on the assumption of the
Milne-Eddington atmosphere \citep[e.g.][]{Lites1990,Klimchuk1992}.

The 180$^{\circ}$ azimuthal ambiguity in the transverse
magnetograms was resolved by using the ``minimum energy" method
\citep{Metcalf1994}. This method uses the simulated annealing algorithm
to minimize a functional $|J_{z}|+|\nabla\cdot\textbf{B}|$, where
the former is the vertical electric current density and the latter
is the field divergence. This method is evaluated to be the
top-performing automated method among state-of-the-art disambiguity
algorithms \citep{Metcalf2006}.

\begin{figure}
\begin{center}
\epsscale{1.1} \plotone{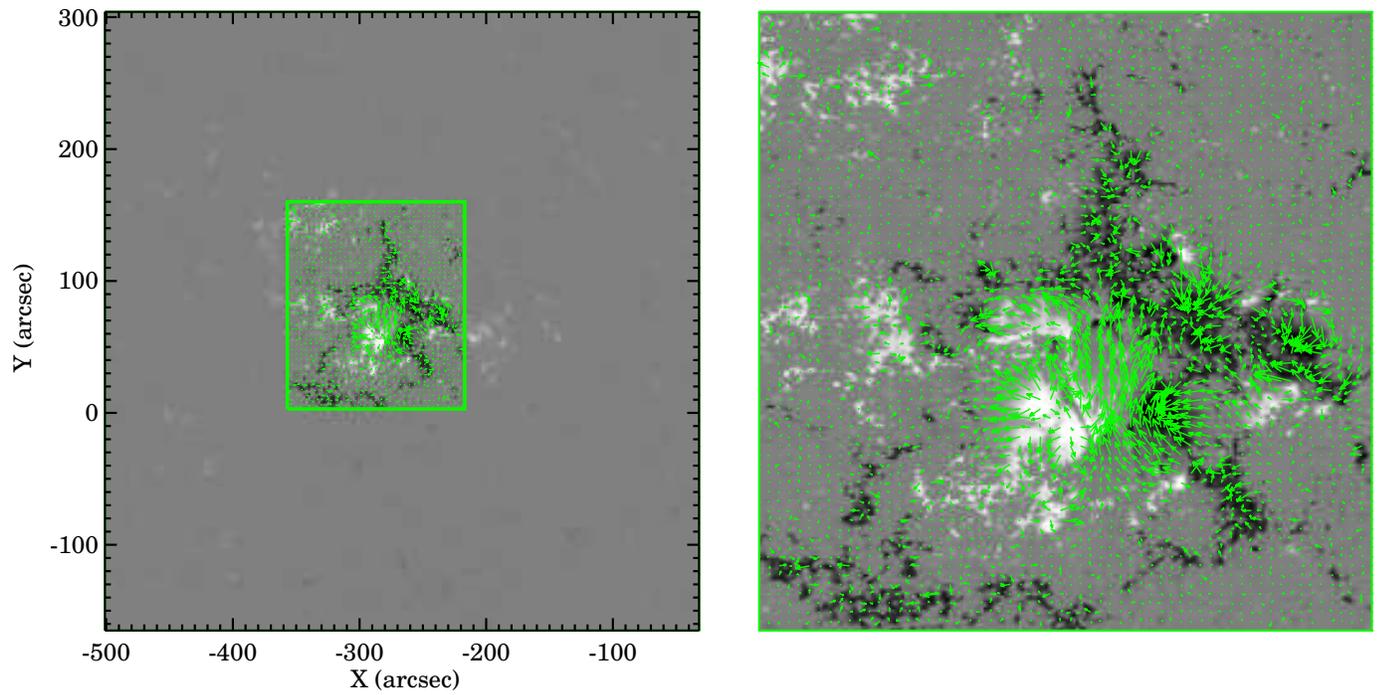} \caption{ \label{magnetogram}Left: the
expanded LOS magnetogram. The green box marks the FOV of the
\emph{Hinode}/SOT-SP vector magnetogram. Right: the close-up view
of the vector magnetogram. Green arrows indicate the transverse
fields.}
\end{center}
\end{figure}

In order to enlarge the field-of-view (FOV) of the photospheric
boundary and incorporate information on magnetic flux outside the
\emph{Hinode}/SOT-SP vector magnetogram, we embedded the
\emph{Hinode}/SOT-SP map into a larger line-of-sight (LOS)
magnetogram that was obtained at 06:27 UT by the Michelson Doppler
Imager (MDI) on board the Solar and Heliospheric Observatory
(\emph{SOHO}). The dimensions of the expanded FOV are 471\arcsec
$\times$ 471\arcsec\ (or 341 Mm $\times$ 341 Mm). The left panel in
Figure~\ref{magnetogram} shows the expanded LOS magnetogram. The green box marks
the FOV of the \emph{Hinode}/SOT-SP vector magnetogram. The right
panel is the close-up view of the vector magnetogram.

Since the photospheric magnetic field does not necessarily satisfy
the force-free condition, we preprocessed the photospheric
boundary using a preprocessing method developed by Wiegelmann et
al. (2006). This preprocessing routine minimizes a functional
$L_{prep}= \mu_1L_1 + \mu_2L_2 + \mu_3L_3 + \mu_4L_4$, where $L_1$
and $L_2$ terms contain force-free and torque-free consistency
integrals, the $L_3$ term controls noise-level, and the $L_4$ term
controls the smoothing. As a result, the forces and torques are
minimized and the preprocessed photospheric boundary is closer to
the force-free condition \citep{Metcalf2008}.

We then applied the weighted optimization method \citep{Wiegelmann2004}
to the preprocessed photospheric boundary to extrapolate the NLFF
field. The weighted optimization method is a modification of the
original optimization algorithm \citep{Wheatland2000}. It
involves minimizing a weight functional of the Lorentz force and
the divergence-free condition throughout the computational domain.
The weighting functions for the force and divergence terms are
position-dependent, with the center of the computational domain
having a weight of 1 and dropping to 0 monotonically in a
buffer boundary region that consists of 32 grid points towards the
side and top boundaries.

Using Wiegelmann's code based on the Green's function method \citep[Schmidt H.~U., in][p.107]{Schmidt1964}, the same line of sight magnetogram has been used to produce a PF extrapolation resulting in a datacube having the same size and resolution as the NLFFF extrapolation.

\section{Comparison of the estimated radio source heights with the NLFFF and PF models}

\begin{figure}
\epsscale{0.8}\plotone{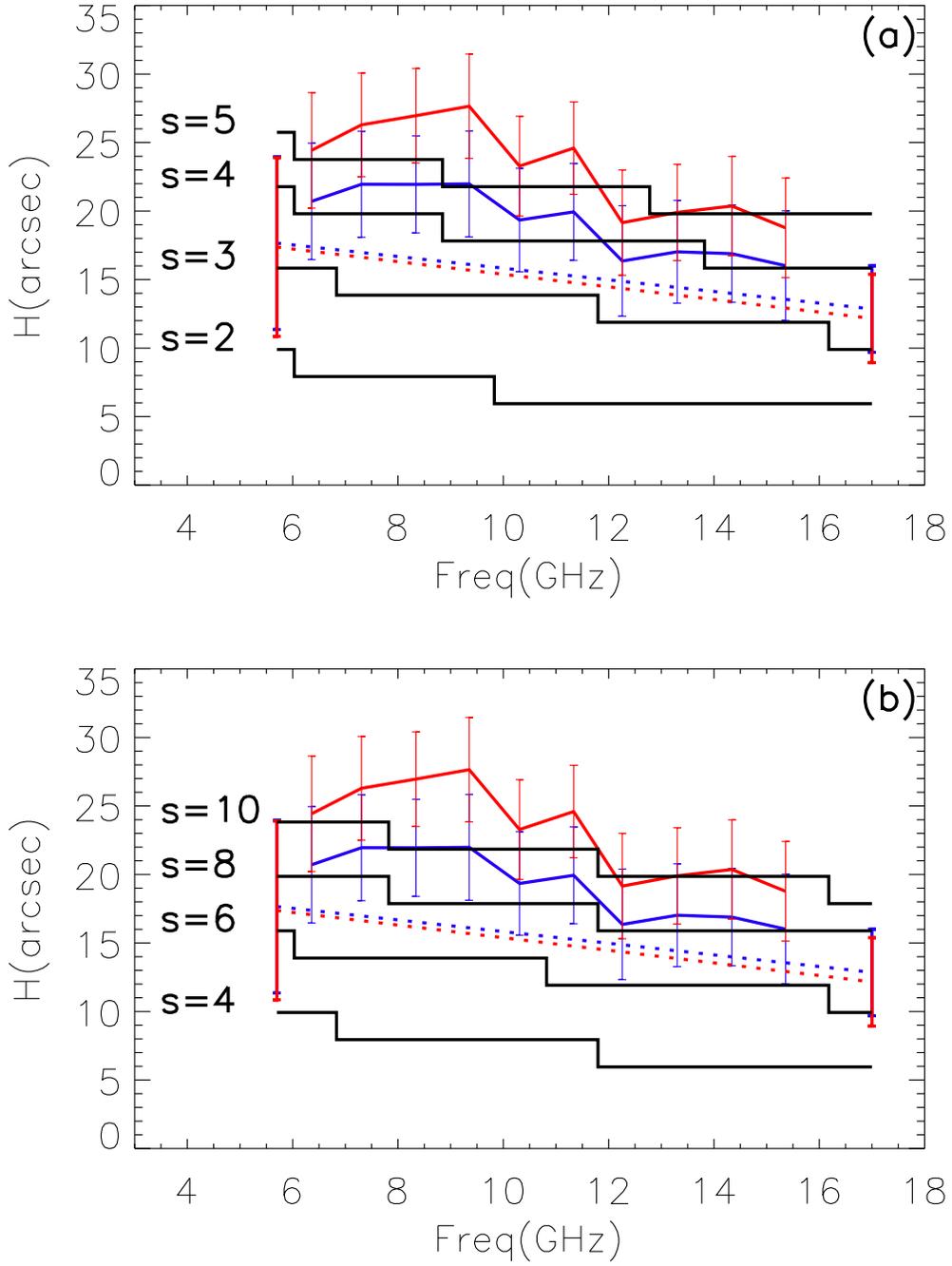}\caption{\label{gyro} ($a$) The frequency dependence of highest heights (black solid lines) where the harmonics $s=2$-5 of the gyrofrequency are available in the NLFFF data cube, superimposed on the radio source height estimates displayed in Figure \ref{h}. ($b$) The frequency dependence of highest heights (black solid lines) where the harmonics $s=4$-10 are available in the PF data cube, superimposed on the radio source height estimates displayed in Figure \ref{h}.}
\end{figure}

The standard GR theory \citep[see, e.g.,][for a review]{Lee_2007} suggests that the bulk of the observed gyroemission comes from the level of the third harmonic ($s=3$) of the gyrofrequency (i.e., $f\approx3f_{Be}$), although some emission can come from the second gyro layer. From this perspective, we first check if there are the required gyro layers at the heights of the radio sources at various frequencies. To do so we mark in Figure~\ref{gyro} the highest heights where the harmonics $s=2$-5 are available in the NLFFF data cube (Fig.~\ref{gyro}$a$) and harmonics $s=4$-10 are available in the PF data cube (Fig.~\ref{gyro}$b$). Inspection of this figure shows that the PF model matches radiation at harmonics $s>6$, while the NLFFF model yields larger coronal field and so is consistent with gyroemission at smaller harmonics. Nevertheless, the mean heights of the observed radio sources is higher, at all frequencies, than the corresponding $s=3$ level, although within the error bars this position can be marginally consistent with $s=3$ at the lowest frequencies. This overall check lacks positional information, since we do not have reliable $x$ and $y$ coordinates for the frequencies provided by RATAN-600.

\begin{figure}
 \centering
 \begin{tabular}{cc}
 \epsfig{file=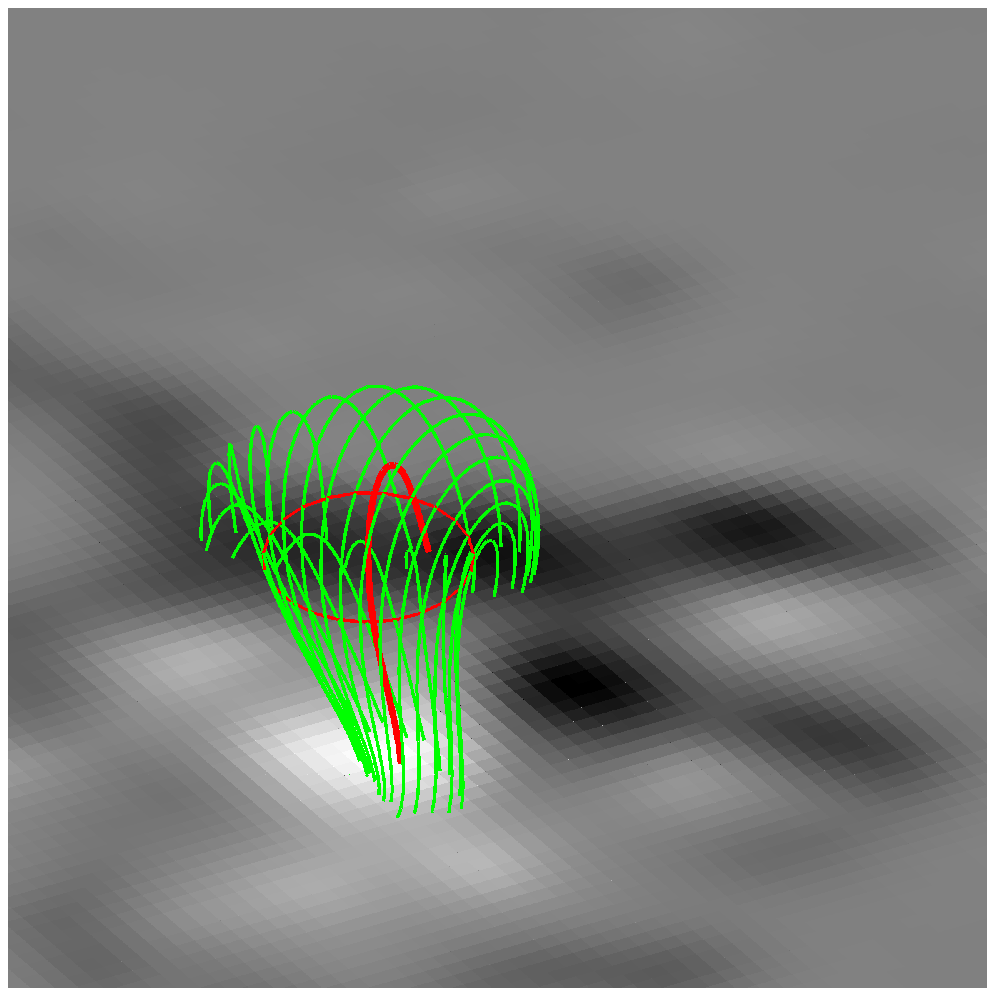,width=0.4\linewidth,clip=} &
 \epsfig{file=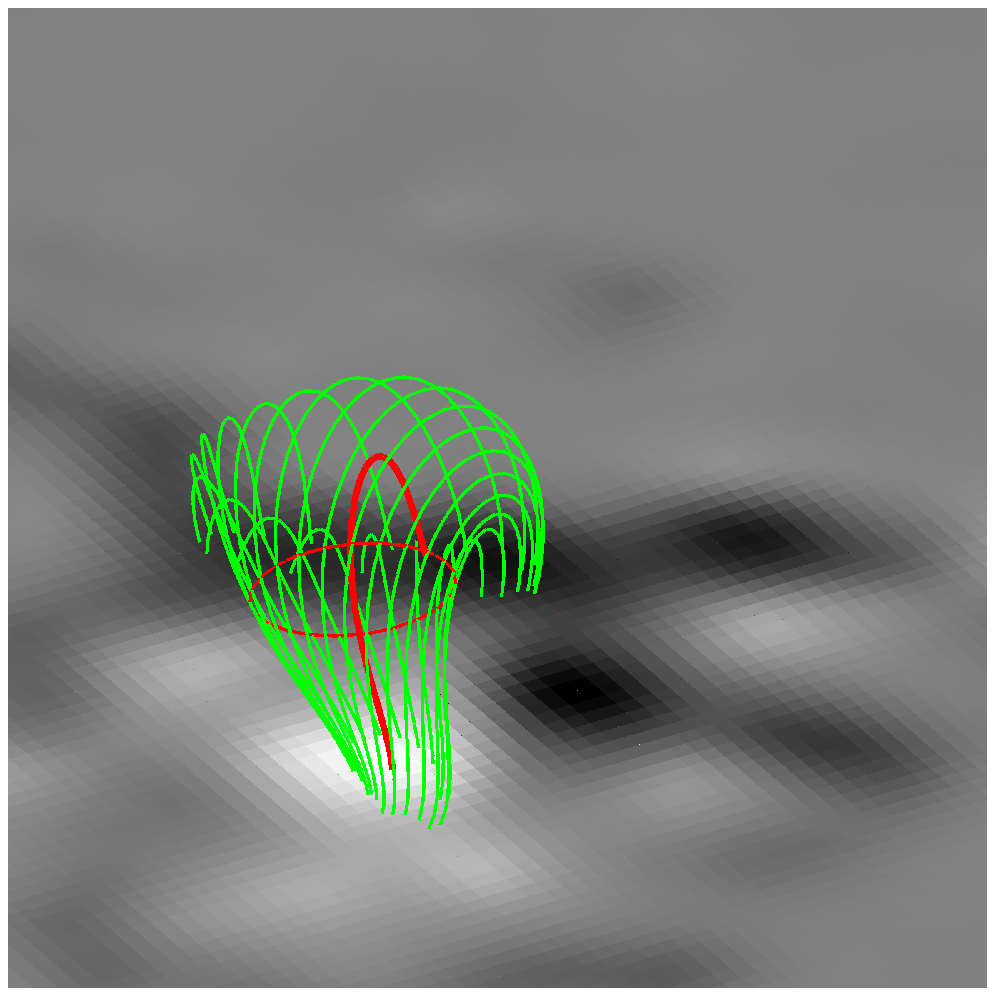,width=0.4\linewidth,clip=} \\
 \epsfig{file=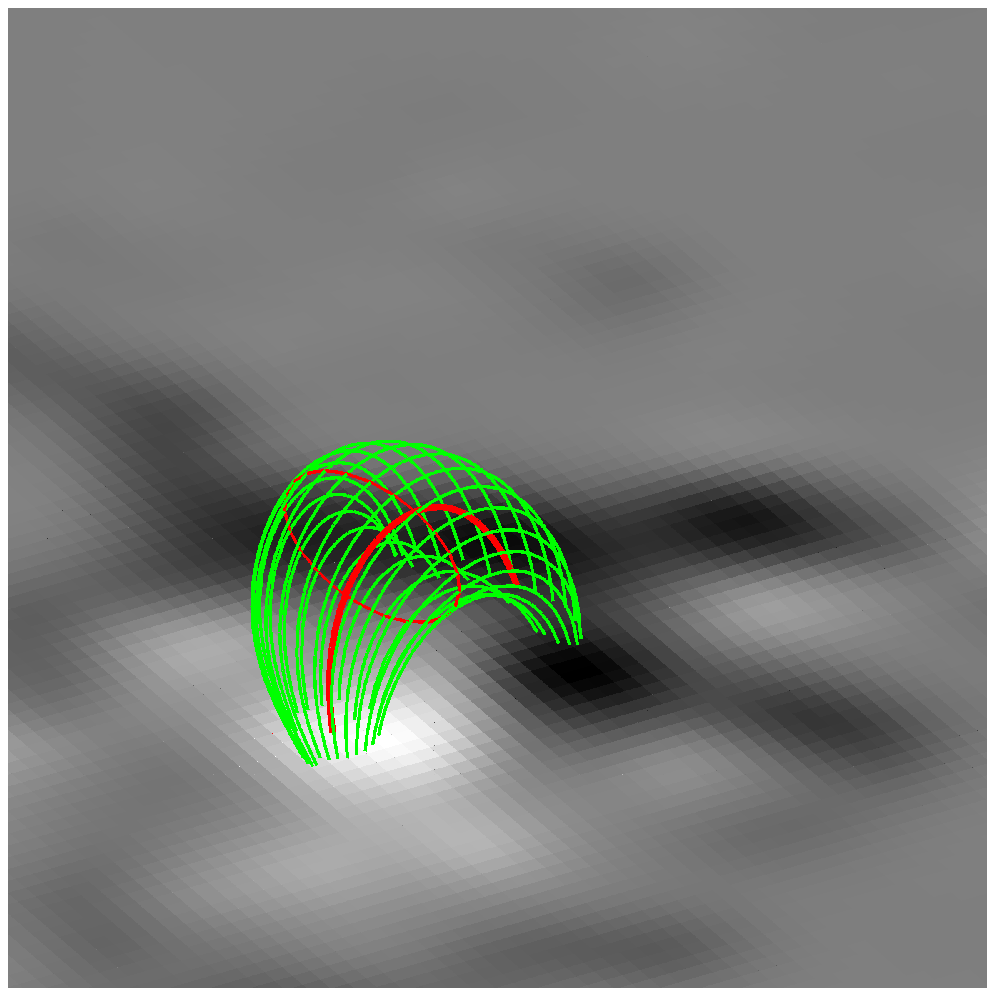,width=0.4\linewidth,clip=} &
 \epsfig{file=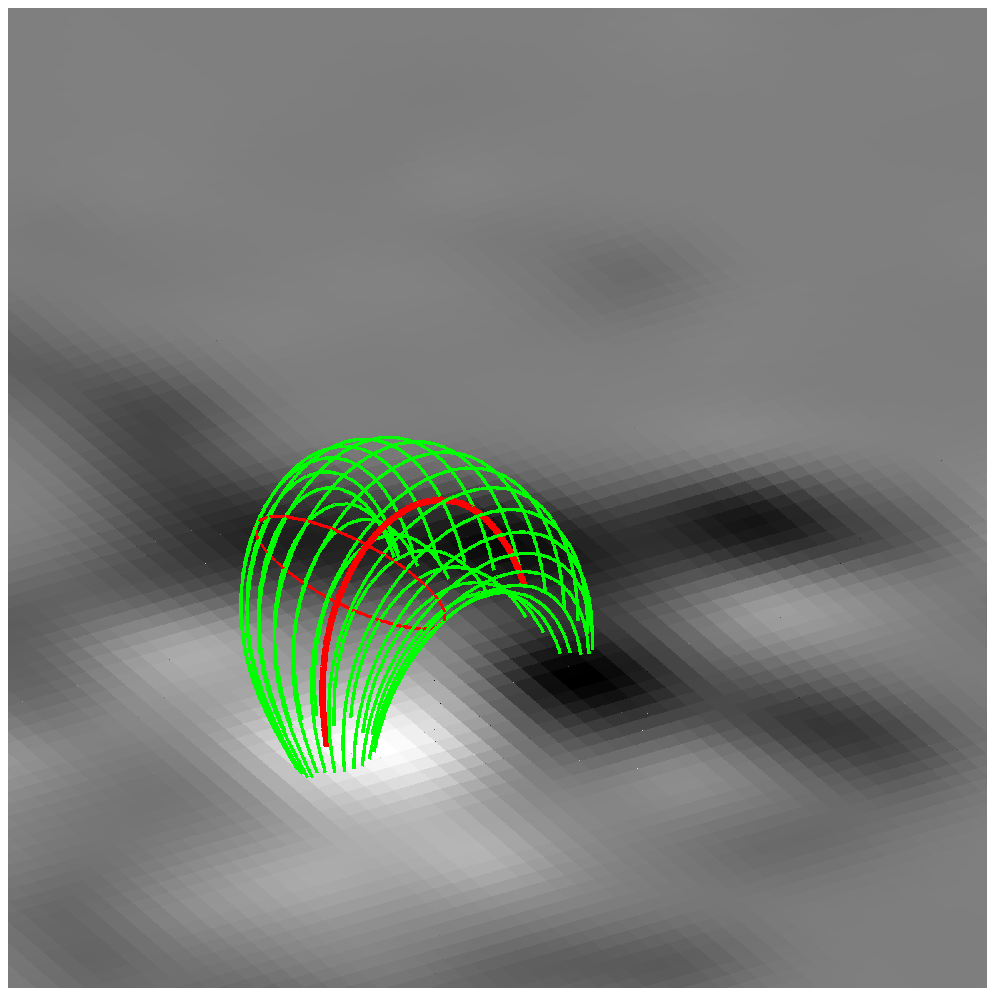,width=0.4\linewidth,clip=}
 \end{tabular}
 \caption{\label{ssrt_perspective} Arbitrary perspective closeup view of the NLFFF (top row) and PF (bottom row) magnetic flux tubes (green lines) corresponding to circular cross sections ($\sim 10"$radius) normal to the field lines (red) passing through the estimated centroids of the LCP SSRT (left column) and RCP SSRT (right column) radio sources at 5.7~GHz. The local coordinates of LCP and RCP centroids derived by solar rotation stereoscopy, $[x=-18.57'', y=-7.82'',z=17.66'']$, and $[x=-22.01'', y=-10.47'', z=17.37'']$, respectively, are relative to center pixel of the line of sight magnetogram map (grey scale), which corresponds to the heliospheric coordinates $[h=0, \phi=-16.199^{\circ}EW,\lambda=1.995^{\circ}NS]$. }
 \end{figure}

\begin{figure}
 \centering
 \begin{tabular}{cc}
 \epsfig{file=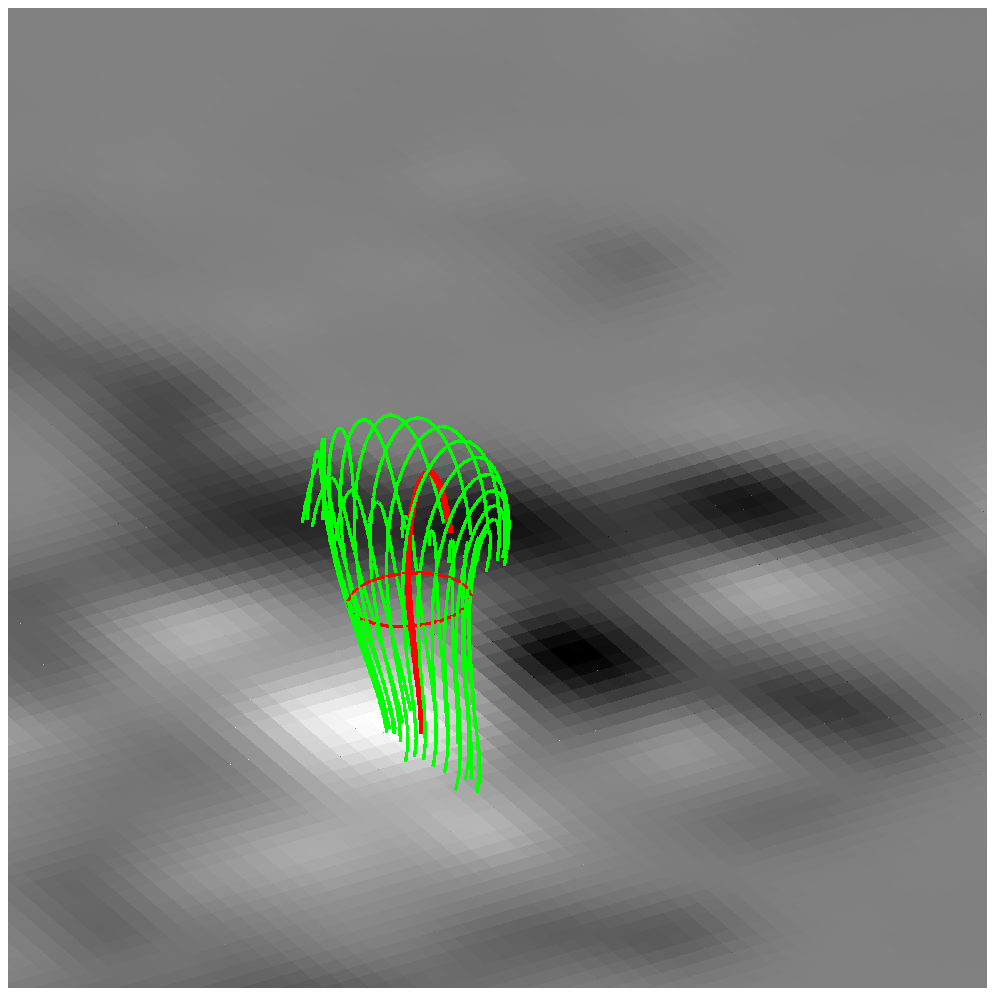,width=0.4\linewidth,clip=} &
 \epsfig{file=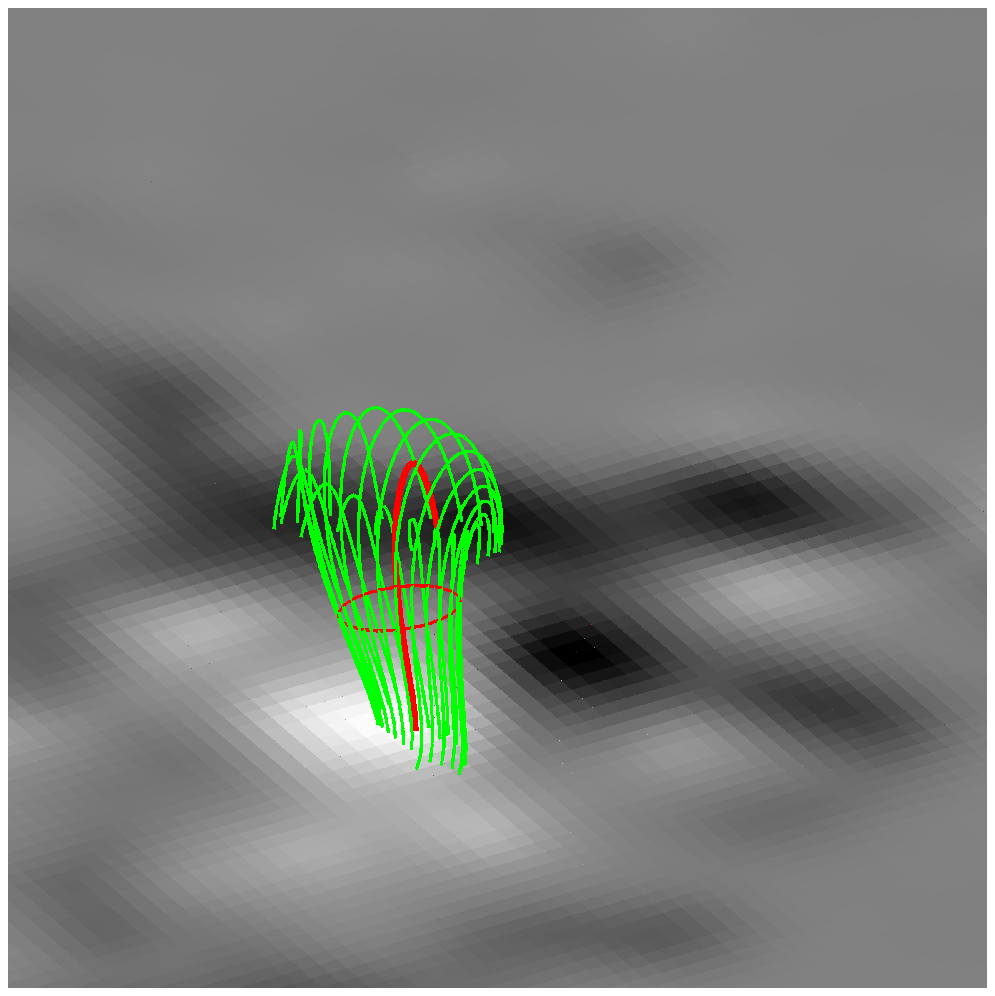,width=0.4\linewidth,clip=} \\
 \epsfig{file=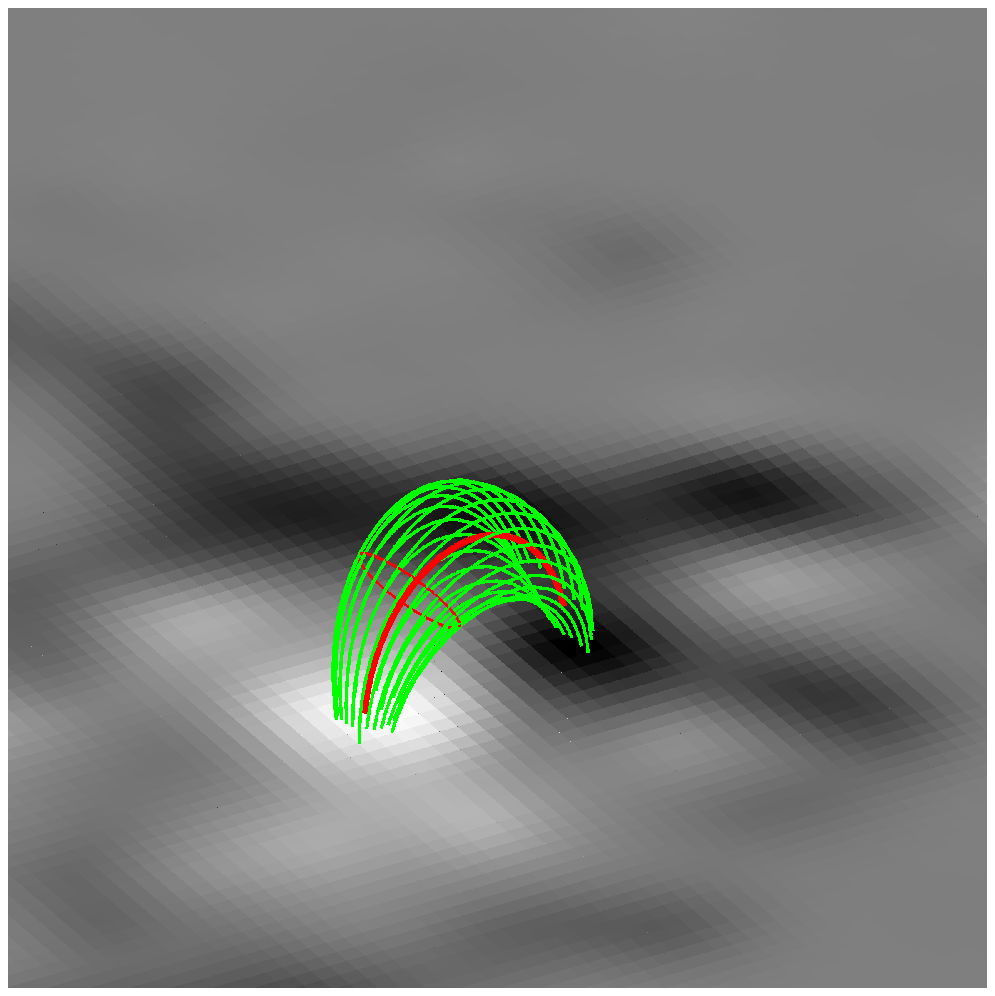,width=0.4\linewidth,clip=} &
 \epsfig{file=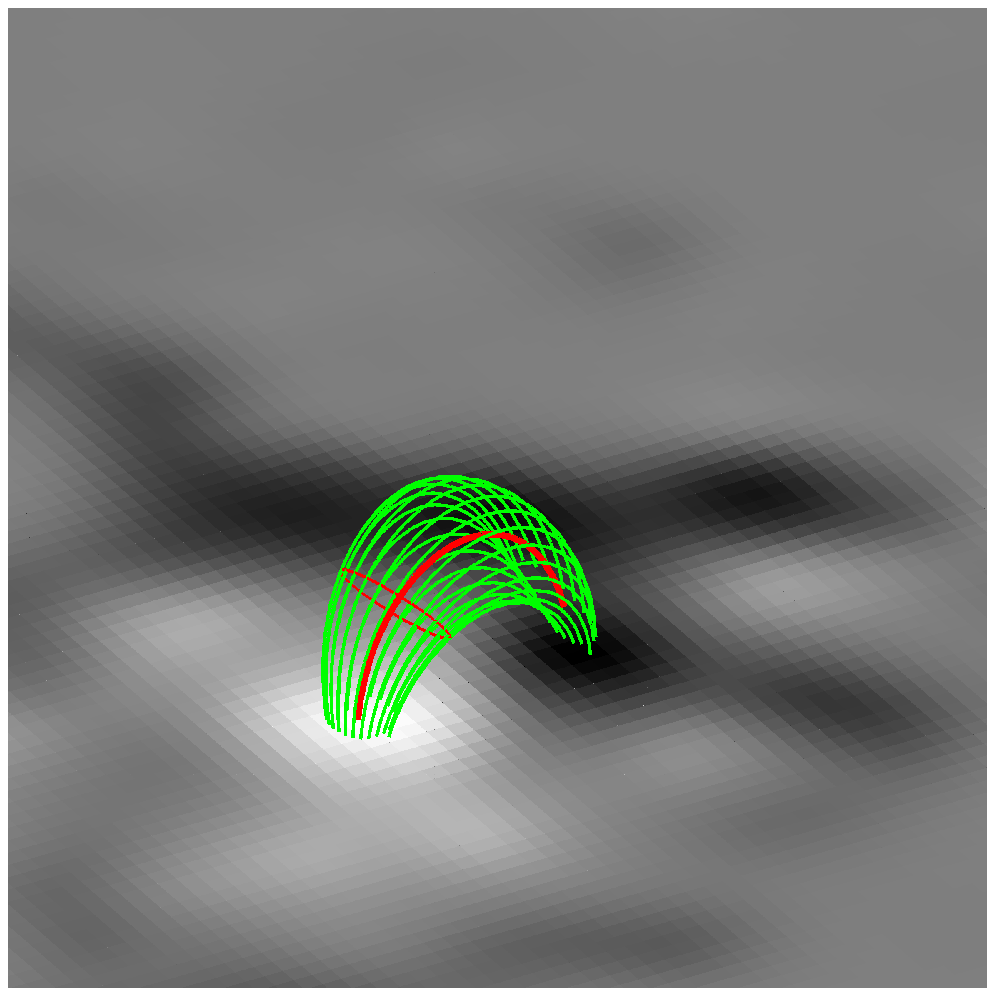,width=0.4\linewidth,clip=}
 \end{tabular}
 \caption{\label{norh_perspective}Arbitrary perspective closeup view of the NLFFF (top row) and PF (bottom row) magnetic flux tubes (green lines) corresponding to circular cross sections ($\sim 6''$radius) normal to the field lines (red) passing through the estimated centroids of the LCP NoRH (left column) and RCP NoRH (right column) radio sources at 17~GHz. The local coordinates of LCP and RCP centroids derived by solar rotation stereoscopy, $[x=-18.04'', y=-14.22'',z=12.85'']$, and $[x=-19.11'', y=-14.16'', z=12.16'']$, respectively, are relative to center pixel of the line of sight magnetogram map (grey scale), which corresponds to the heliospheric coordinates $[h=0, \phi=-16.199^{\circ}EW,\lambda=1.995^{\circ}NS]$. }
 \end{figure}

To make a meaningful comparison between the 3D datacubes of the extrapolated magnetic fields and 3D positions of the radio sources determined from the imaging data, we have developed a visualization tool capable of importing the 3D magnetic model, computing and plotting a field line passing through an arbitrary point within this cube, building a flux tube around this field line, and inspecting the magnetic field/isogauss surfaces. Figure \ref{ssrt_perspective} displays, from a single arbitrary direction, a set of closeup views of the magnetic flux tubes built around the central field line passing through the derived height of the 5.7~GHz SSRT LCP (left column) and RCP (right column) radio sources, as obtained from the NLFFF (top row) and PF (bottom row) extrapolations. Having the same 4-panel structure, Figure \ref{norh_perspective} displays a set of flux tubes corresponding to the estimated positions of the 17~GHz NoRH LCP and RCP radio sources.

\begin{figure}
\epsscale{1.1}\plotone{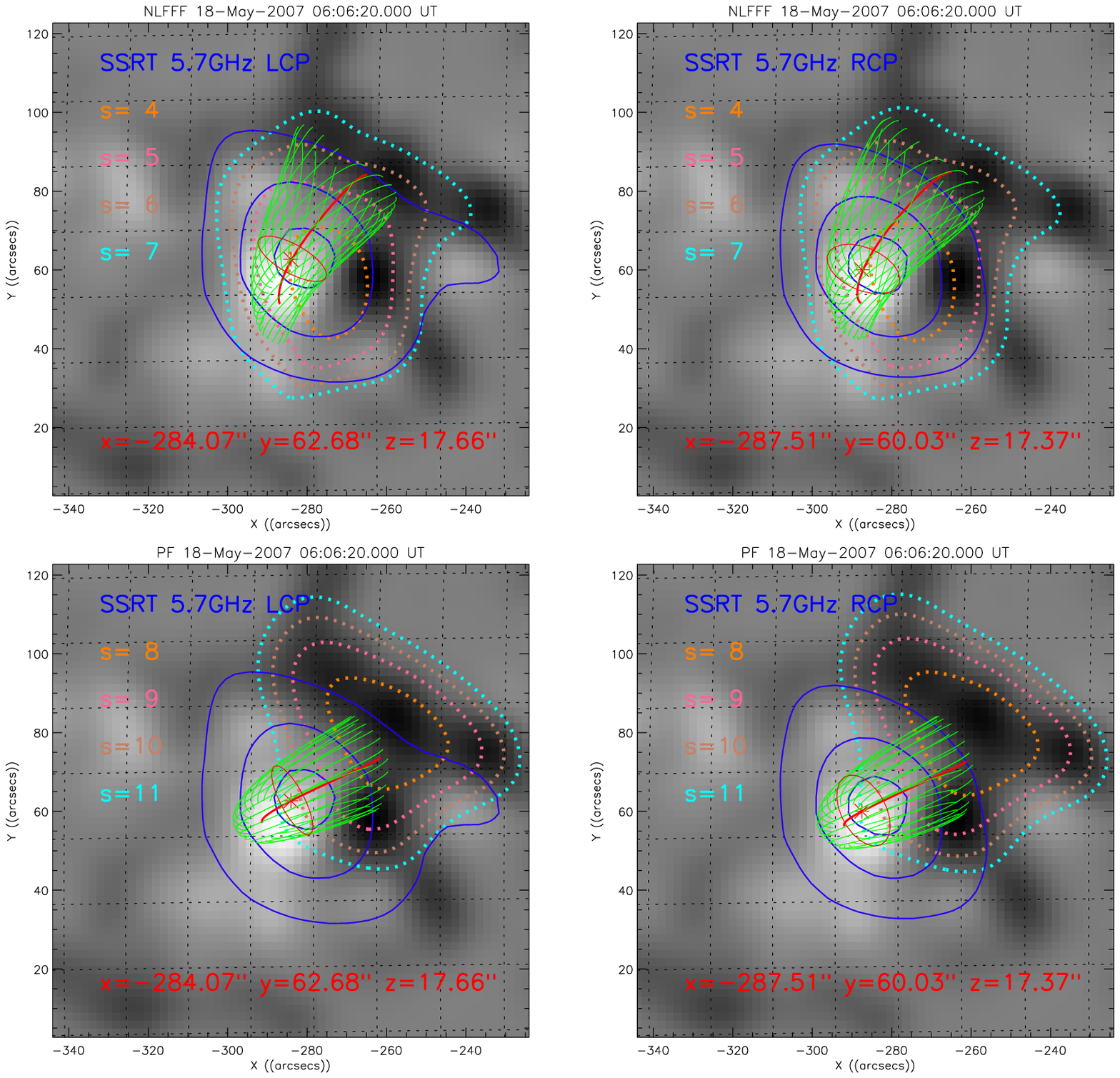}
\caption{\label{ssrt_maps} Line of sight view of the the same magnetic flux tubes (green lines) shown in Figure \ref{ssrt_perspective}. The central field lines and normal circular cross sections defining each flux tubes are shown in red. The estimated positions of the SSRT radio sources are indicated by red asterisk symbols, and their heliocentric coordinates, measured in arcseconds, are indicated on each map. The solid blue lines indicate the $15\%$, $50\%$, and $85\%$ radio contours corresponding to the LCP (left column) and RCP (right column) polarizations. The color coded dotted lines indicate the first four existing isogauss contours corresponding to the 5.7~GHz gyroresonance harmonics indicated in each legend, as derived from the NLFFF (top row) and PF (bottom row) extrapolations at the estimated radio source heights.    }
\end{figure}

\begin{figure}
\epsscale{1.1}\plotone{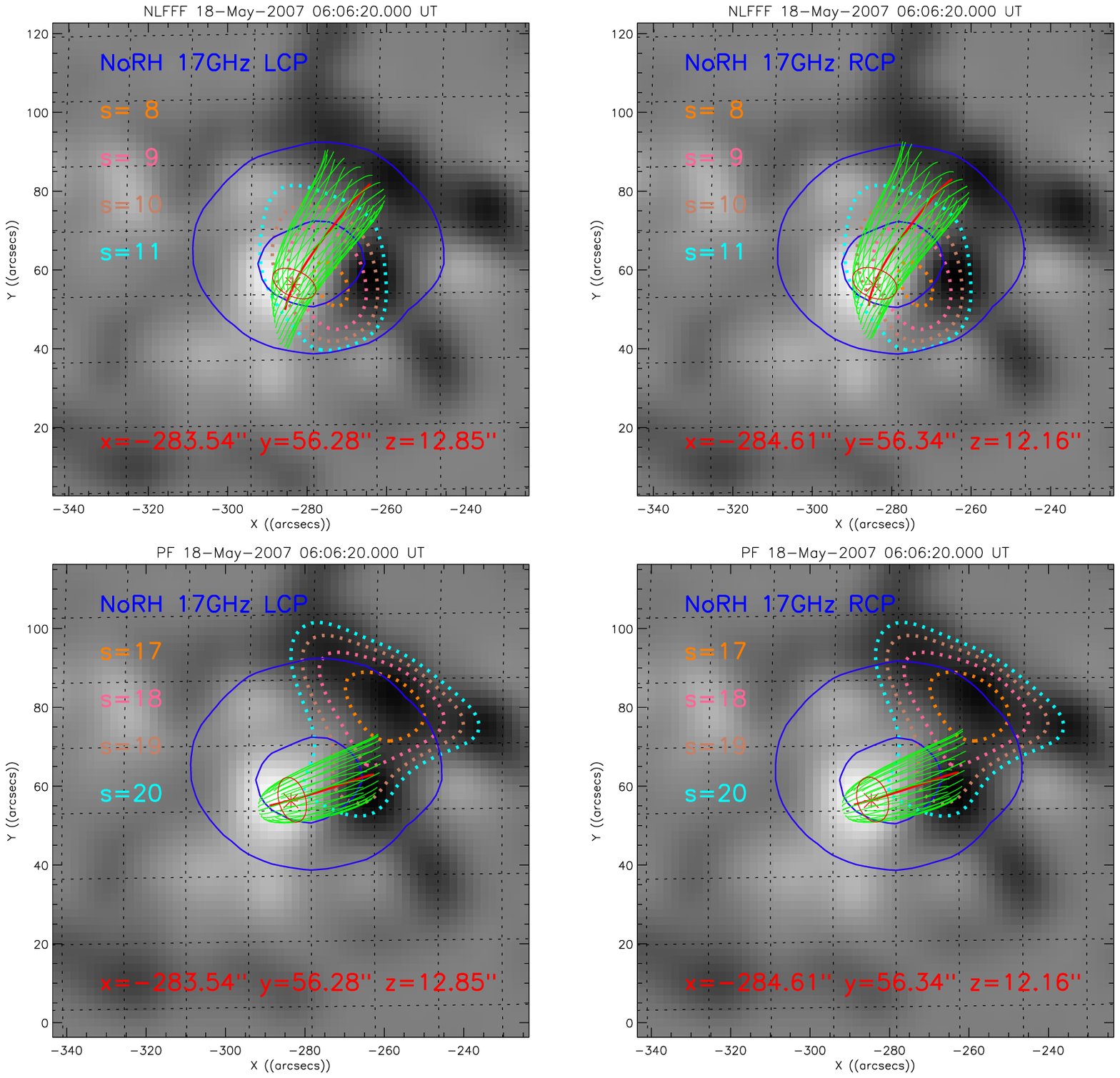}
\caption{\label{norh_maps} Line of sight view of the the same magnetic flux tubes (green lines) shown in Figure \ref{norh_perspective}. The central field lines and normal circular cross sections defining each flux tubes are shown in red. The estimated positions of the NoRH radio sources are indicated by red asterisk symbols, and their heliocentric coordinates, measured in arcseconds, are indicated on each map. The solid blue lines indicate the $15\%$, $50\%$, and $85\%$ radio contours corresponding to the LCP(left column) and RCP(right column) polarizations. The color coded dotted lines indicate the first four existing isogauss contours corresponding to the 17~GHz gyroresonance harmonics indicated in each legend, as derived from the NLFFF(top row) and PF(bottom row) extrapolations at the estimated radio source heights.  }
\end{figure}

Figure~\ref{ssrt_maps} displays, from the Earth-Sun line of sight perspective, the results of the NLFFF-PF comparison for the SSRT source. We see that closed magnetic field lines (red solid lines) corresponding to the derived radio source positions (asterisk symbols) cross the observed radio centroid positions, although these lines differ in the two cases. Moreover, in case of the PF extrapolation, the isogauss contours (dotted lines) at the lowest available gyro harmonics (8th to 11th) at the radio source height level are displaced from the radio source position, the lowest harmonics available at the radio source position being $s=11$. The picture is essentially different in the case of the NLFFF extrapolation: the isogauss lines are much better co-aligned with the radio contours, there are 4th and 5th harmonics layers at and around the radio source location, but there is no $s=3$ layer. A similar comparison with the NoRH radio sources (Figure~\ref{norh_maps}) shows that the NLFFF model implies the radio source at $s=8$-9, while  in the case of the potential field model the radio source height corresponds to $s>20$, far higher than $s=3$ for both models. No systematic error in the height estimate can change this conclusion: the third harmonic resonance for 17~GHz (2.0 kG) is unavailable in the entire extrapolation data volume in both models, even at their bottom layer.

\section{Discussion}

We have found that the determined mean positions of the radio centroids are located higher in the corona than the corresponding 3rd harmonic gyroresonance layer at all available frequencies and for both models of coronal magnetic field. For the PF model this mismatch is extreme and, in addition, the isogauss contours are significantly displaced from the apparent radio source position. We must conclude that the PF model does not offer a quantitatively good description of the coronal magnetic field for this particular AR.

For the NLFFF model there is a good match between the radio brightness and isogauss contours, which makes the NLFFF model more appropriate for our case. Nevertheless, there is still a mismatch, also reported for some other events \citep[e.g.,][]{Uralov_etal_2006, Uralov_etal_2008}, between the standard GR theory prediction of the emission at $s=2$-3 and observed emission from a region of higher harmonics, which is addressed below. We consider three possible reasons for this mismatch: instrumental/algorithm errors, errors in either the assumed coronal temperature or NLFF field extrapolation, and inadequacy of the theory (which involves magnetic extrapolation and GR theory).

Indeed, one can argue that the third gyro layer is consistent within the error bars with the source position at the lowest frequencies. However, is does not seem likely because the direct SSRT imaging of the AR above the limb confirms the height estimate.

The second option, the extreme parameter range, could in fact work, but would require a high plasma temperature ($\sim10^7$~K) leading to enhanced optical depth of the gyroemission at the fourth harmonic of the gyrofrequency. However, the measured brightness temperature of the radio emission (Figure~\ref{h}$b$) does not favor this option; the temperature is well below 3~MK, which is insufficient for the fourth harmonic to play a role. There are also no context observations that would imply such a high temperature at this region. We note that the brightness temperature in Figure~\ref{h}$b$ decreases with frequency, so the emission above $\sim12$~GHz can contain a significant free-free component, along with a (possibly variable) GR component.  We cannot exclude an error in the NLFF field extrapolation. Indeed, the extrapolations of the magnetic field from the photospheric level to the corona are known to suffer from a fundamental problem of the need to extrapolate from the forced photospheric boundary to the force-free corona, which requires data pre-processing that necessarily introduces an uncertainty to the extrapolated field. In addition, in our case we have only a limited field of view for the vector magnetogram, while for the remainder of the field of view the line-of-sight magnetogram was used, which is another source of uncertainty. Unfortunately, there is currently no independent way of verifying the magnetic field model, although it might become possible in some cases with infrared coronal magnetography and with other radio methods.

Finally, we envision a possible shortcoming of the GR theory. The commonly used formulae of the GR optical depth are obtained within an assumption of a steady-state Maxwellian distribution of the radiating electrons, which has never been confirmed (it must be noted that the corona is an open, rather than isolated, system driven by a sustained external energy flux). In a similar open system, the solar wind, which is in fact an expanding part of the solar corona, in situ measurements never yield the Maxwellian distribution, but rather better fit a so-called kappa-distribution (roughly speaking a Maxwellian core with a power-law tail with different indices). If this or a similar (possibly time-varying) distribution is also typical for the lower corona, from which we observe the GR radio emission, the GR optical depth at the harmonics higher than $s=3$ will be greatly enhanced by electrons in the nonthermal tail of the distribution, in possible agreement with the measured heights of the radio sources. This option is highly suggested by our detection of gyro emission at up to 17~GHz, where the mismatch with the third gyro harmonic is especially clear.

\section{Conclusions}

We have reported multi-instrument, multifrequency rotation stereoscopy of an active region. We found that for this particular case the NLFFF represents a much better field model than the PF. The remaining mismatch between the magnetic field structure, radio source heights, and GR theory raises fundamental questions in coronal plasma physics, including adequacy of the magnetic extrapolations and validity of the assumption of a Maxwellian distribution of the coronal plasma. The latter question, i.e., the question of what is the true distribution function of an open system composed of tenuous plasma, along with the closely related, yet unanswered question of the mechanisms of coronal heating, is highly important for solar physics and, indeed, for collisionless plasma astrophysics in general.  Comparison of AR radio emission with NLFFF extrapolations is a fruitful way of addressing both questions.

\acknowledgments
This work was partly supported by the NASA NNX08-AJ23G, NNX08-AQ90G, and NNX10AF27G grants and NSF grants
AGS-0961867 and AST-0908344 to the New Jersey Institute of Technology, by the Russian Academy os Sciences programs OFN-16 and PAN-16, and by the Russian Foundation for Basic Research grants 08-02-00378, 09-02-00111, 09-02-00226, 09-02-00624, and 11-02-91175.

\bibliographystyle{apj}
\bibliography{AR_bib,WP_bib} 

\begin{deluxetable}{lccccc}
\tablecaption{Derived parameters versus exact solution 
\label{test}}
\tablehead{ \colhead{Method} &\colhead{$h/R_{Sun}$} &\colhead{$\delta\varphi$} &\colhead{$\delta\lambda$}  &\colhead{$\sigma$} &\colhead{$\sigma/R_{Sun}^*$}\\&(\%)&(deg)&(deg)&(arcsecs)&(\%)}
\startdata
Exact Solution& 1.000&-0.500& 0.500& 1.930& 0.203\\
2D($ \rho, \delta\varphi, \delta\lambda$)& 0.955&-0.523& 0.497& 1.917& 0.202\\
1D($ \rho, \overline{\delta\varphi}, \overline{\delta\lambda}$)&
 0.963&-0.523& 0.497& 1.858& 0.196\\
1D($ \rho, \delta\varphi, \overline{\delta\lambda}$)&
 0.965&-0.521& 0.497& 1.863& 0.196\\
1D($ \rho, \overline{\delta\varphi}, \delta\lambda$)&
 0.967&-0.523& 0.502& 1.863& 0.196\\
1D($\rho, \delta\varphi, \delta\lambda$)& 0.984&-0.544& 0.558& 1.868& 0.197
\enddata
\end{deluxetable}

\end{document}